\documentclass[prl,twocolumn,showpacs,superscriptaddress]{revtex4}

\usepackage[dvips]{graphicx}
\usepackage{amsmath}
\usepackage{color}

\newcommand{\etal}{\textit{et al.~}}

\begin{document}

\title{Dynamical Dzyaloshinsky-Moriya Interaction in KCuF$_3$}

\author{M.~V.~Eremin} \affiliation{EP V, Center for Electronic
Correlations and Magnetism, University of Augsburg, 86135 Augsburg,
Germany}

\affiliation{Kazan State University, 420008 Kazan, Russia}

\author{D.~V.~Zakharov*}

\affiliation{EP V, Center for Electronic Correlations and Magnetism,
University of Augsburg, 86135 Augsburg, Germany}

\author{H.-A.~Krug~von~Nidda}

\affiliation{EP V, Center for Electronic Correlations and Magnetism,
University of Augsburg, 86135 Augsburg, Germany}

\author{R.~M.~Eremina}
\affiliation{EP V, Center for Electronic Correlations and Magnetism,
University of Augsburg, 86135 Augsburg, Germany}

\affiliation{E. K. Zavoisky Physical Technical Institute, 420029
Kazan, Russia}

\author{A.~Shuvaev}
\affiliation{EP 4, Physikalisches Institut, Universit\"{a}t
W\"{u}rzburg, D-97074 W\"{u}rzburg, Germany}

\author{A.~Pimenov}
\affiliation{EP 4, Physikalisches Institut, Universit\"{a}t
W\"{u}rzburg, D-97074 W\"{u}rzburg, Germany}

\author{P.~Ghigna}
\affiliation{Dipartimento di Chimica Fisica, Universit\`{a} di
Pavia, I-27100 Pavia, Italy}

\author{J.~Deisenhofer}
\affiliation{EP V, Center for Electronic Correlations and Magnetism,
University of Augsburg, 86135 Augsburg, Germany}

\author{A.~Loidl}
\affiliation{EP V, Center for Electronic Correlations and Magnetism,
University of Augsburg, 86135 Augsburg, Germany}

\date{\today}

\begin{abstract}
The spin dynamics of the prototypical quasi one-dimensional
antiferromagnetic Heisenberg spin $S=1/2$ chain KCuF$_3$ is
investigated by electron spin resonance spectroscopy. Our analysis
shows that the peculiarities of the spin dynamics require a new
\textit{dynamical} form of the antisymmetric anisotropic spin-spin
interaction. This dynamical Dzyaloshinsky-Moriya interaction is
related to strong oscillations of the bridging fluorine ions
perpendicular to the crystallographic $c$ axis. This new mechanism
allows to resolve consistently the controversies in observation of
the magnetic and structural properties of this orbitally ordered
perovskite compound.

\end{abstract}


\maketitle



The pervoskite system KCuF$_3$ is presumably one of the best
realizations of an one-dimensional (1D) antiferromagnetic (AFM)
Heisenberg chain as a direct consequence of the orbital ordering in
this compound \cite{Kugel82}. Recently, detailed inelastic neutron
scattering investigations have shown that fingerprints of a 1D
Luttinger liquid, namely the spinon excitation continuum, exist up
to temperatures of about 200 K \cite{Lake05}. Moreover, these
specifically 1D features are still observable below the onset of
antiferromagnetic ordering at $T_{\rm N} = 39$~K, indicating the
strong quantum nature of magnetism in KCuF$_3$.

\begin{figure}[b]
\centering
\includegraphics[width=85mm]{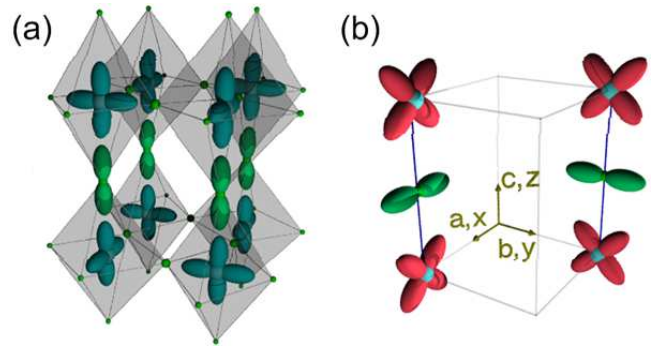}
\caption{(color online) (a): Schematic orbital ordering in the
ground state of type-$a$ KCuF$_3$. The big (small) spheres denote Cu
(F) ions. The K ions, placed between the CuF$_6$ octahedra, are not
shown. (b): Two most relevant exchange paths of the ring-like SAE
between the excited states of Cu ions.} \label{SAEorb}
\end{figure}

In spite of its paradigmatic status, the driving forces of the
orbital and magnetic structure are far from being understood at the
moment. Even the crystal structure originally assigned to be
tetragonal $D_{4h}^{18}$ (see Fig.~\ref{SAEorb}) was claimed
\cite{Hidaka98} to be orthorhombic $D_2^4$. Electron spin resonance
(ESR) measurements \cite{Yamada89,Yamada94} played a key role in
triggering these investigations: they suggested the existence of the
antisymmetric anisotropic exchange, usually termed as
Dzyaloshinsky-Moriya (DM) interaction, thereby questioning the
crystal symmetry determined earlier. But the orthorhombic distortion
deduced from x-ray diffraction \cite{Hidaka98} is not consistent
with recent NQR data \cite{Mazzoli04} and other experimental and
theoretical findings \cite{Yamada94, Binggeli04}.

In this Letter we reanalyze all \emph{static} sources of the ESR
line broadening on the microscopical level and show that the
discrepancies in the understanding of this paradigm compound can be
resolved by introducing a \emph{dynamical} Dzyaloshinsky-Moriya
(dDM) interaction.

\begin{figure}[tbp]
\centering
\includegraphics[width=80mm]{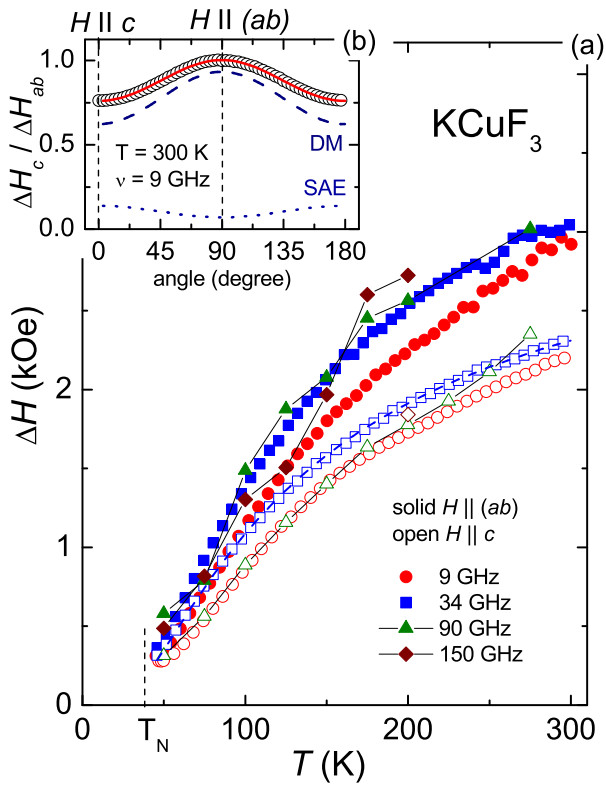}
\caption{(color online) (a) Temperature dependence of the linewidth
at different frequencies. The blue dashed line (34~GHz, $H || c$)
represents a fit to an Arrhenius law. (b) Angular dependence of
$\Delta H$ at $T = 300$~K. The fit is described in the text.}
\label{Exp}
\end{figure}

Single crystals of KCuF$_3$ have been grown as described in
Ref.~\onlinecite{Caciuffo02}. The ESR experiments were carried out
with a Bruker ELEXSYS E500 CW-spectrometer at frequencies of 9 and
34~GHz. For measurements at 90 and 150~GHz a quasi-optical technique
was used \cite{Ivannikov02}. The ESR spectrum consists of a single
exchange narrowed Lorentz line at resonance fields corresponding to
$g_{c} = 2.15$ and $g_{a} = 2.27$ in accordance with previous
experiments at 24~GHz \cite{Ishii90} and 34~GHz \cite{Ikebe71}.
Figure~\ref{Exp} shows the temperature dependence of the ESR
linewidth for different frequencies. Above $T_{\rm N}$ the linewidth
increases monotonously from 0.3~kOe to more than 3 kOe above room
temperature with a pronounced anisotropy with respect to the
$c$-axis (see Fig.~\ref{Exp}(b)), but without any anisotropy within
the $ab$ plane in agreement with the all over tetragonal crystal
structure. For different frequencies the linewidth data follow
approximately the same temperature dependence with comparable
absolute values. It can be described phenomenologically by an
Arrhenius law  $\Delta H \propto \exp(-\Delta/T)$ with an energy gap
$\Delta = 114$~K (see Fig.~\ref{Exp}).

Using conventional estimates \cite{Zakharov08} for all relevant
sources of the ESR linewidth, it was concluded that the relaxation
contribution of the antisymmetric DM interaction is several orders
of magnitude larger than all other mechanisms in KCuF$_3$
\cite{Yamada89}. However, the above mentioned inconsistencies
concerning the direction of the DM vector
\cite{Yamada89,Hidaka98,Yamada94} and a general theoretical analysis
of the spin relaxation in $S=1/2$ chains
\cite{Oshikawa02,Ivanshin03} questioned the validity of the original
approach. Moreover, the microscopic analysis of the exchange paths
in several one-dimensional magnets \cite{KrugvNidda02, Eremina03,
Eremin05,Zakharov05} showed that the symmetric anisotropic exchange
(SAE) may strongly exceed the conventional estimate and thus
dominate the spin relaxation. Therefore, it is necessary to
reanalyze microscopically the magnitude of SAE and DM interactions
in KCuF$_3$.

Let us reconsider the structure and the orbital ordering in
KCuF$_3$. The crystal has a pseudo-perovskite structure with $a =
5.855~${\AA}, $c = 7.852~${\AA}. Each Cu$^{2+}$ ion (electronic
configuration $3d^9$, spin $S=1/2$) is surrounded by a tetragonally
deformed fluorine octahedron. The principal axes of the octahedra
are oriented alternating along the $a$ and $b$ axes, leading to the
orbital ordering of the ground $\vert y^2-z^2 \rangle$ and $\vert
x^2-z^2 \rangle$ states of a hole on Cu ions (see
Fig.~\ref{SAEorb}(a)). The existing small contribution of the $\vert
3z^2 - r^2 \rangle$ orbital to the ground state is not significant
for the anisotropic exchange and will be neglected in the following
analysis.

The crystal structure and the orbital ordering in KCuF$_3$ allow for
SAE interaction ${\cal H}_{\rm SAE}^{(ab)} = S^a_\alpha \cdot
D_{\alpha \beta }^{(ab)} \cdot S^{b}_\beta$ ($\{\alpha, \beta\} =
\{x,y,z\}$, $a,b$ are the interacting ions) only along the
crystallographic $c \equiv z$ axis via the ring-like processes
described in Ref.~\onlinecite{Eremin05}. From the microscopical
point of view, the exchange process consists of the excitation of a
spin via spin-orbit (SO) coupling into an excited state of the
starting ion $a$, the following hopping into an excited orbital
state on the neighboring ion $b$ and its transfer back into the
initial state via the ground orbital state of the ion $b$.
Fig.~\ref{SAEorb} shows that the overlap between the ground and the
relevant excited orbital states ($\vert xz \rangle$, $\vert yz
\rangle$) via F$^-$ $p$ orbitals is of $\sigma$ and $\pi$ type,
respectively. Denoting the respective hopping integrals by
$t'_\sigma$ and $t_\pi$, we can estimate the exchange constant of
SAE like in Ref.~\onlinecite{Eremin05} as
\begin{equation} \label{Destim}
D_{zz} \ll D_{xx} = D_{yy} \approx 4 \lambda^{2} \frac{t_{\pi}
t_{\sigma}'}{\Delta_{cf}^{2} \Delta_{ab}} \approx 1~\textrm{K},
\end{equation}
where $\lambda$ is the SO coupling constant, $\Delta_{cf}$ and
$\Delta_{ab}$ denote the crystal-field splitting and the
charge-transfer energy, respectively. Here we used $\lambda_a /
\Delta_{cf} = 0.05$ \cite{KrugvNidda02} and $4 t_\pi t'_{\sigma} /
\Delta_{ab} \approx 4 t'^2_{\sigma} / \Delta_{ab} \approx 380$~K
\cite{Satija80}.

The obtained value of SAE turns out to be strongly enhanced as
compared to the conventional estimate \cite{Yamada94,Satija80}, but
the resulting ESR line broadening with $\Delta H \sim 10^2$~Oe
cannot explain the huge ESR linewidth $\Delta H \sim 3$~kOe observed
in KCuF$_3$. Note that in all compounds where SAE plays a dominant
role \cite{KrugvNidda02,Eremina03,Eremin05} the characteristic
linewidth at $T \approx J / k_{\rm B}$ is about hundreds of Oe,
i.~e. an order of magnitude smaller than in the case of KCuF$_3$.
This large discrepancy indicates the existence of an additional
source of line broadening in KCuF$_3$. Indeed, Oshikawa and Affleck
\cite{Oshikawa02} supposed another type of spin relaxation to be
dominant in this system, because KCuF$_3$ does not obey the
universal scaling behavior of $\Delta H(T/J)$ typical for spin
relaxation via SAE. Therefore, we will turn now to the analysis of
the second possible source of the ESR line broadening in KCuF$_3$ --
the DM interaction:

\begin{figure}[bp]
\centering
\includegraphics[width=\linewidth]{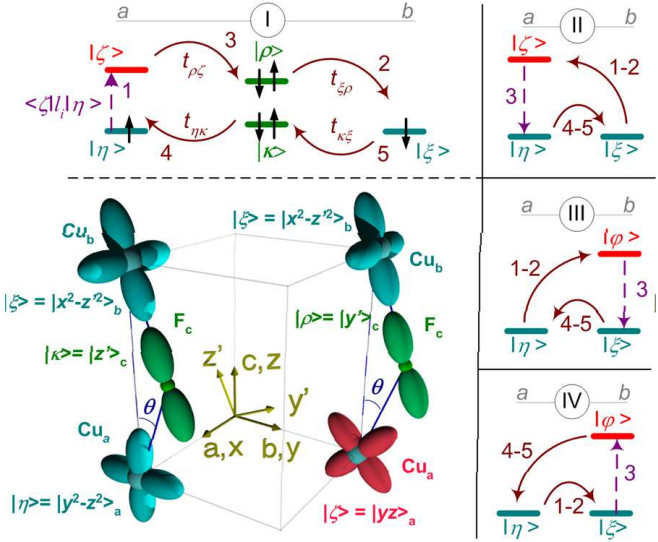}
\caption{(color online) Possible paths for DM interaction between
two sites $a$ (with ground state $\eta$, exited state $\zeta$) and
$b$ ($\xi$ and $\varphi$, respectively). Solid arrows correspond to
the effective hopping integrals, dashed arrows indicate the matrix
elements of the spin-orbit coupling. Numeration indicates one of the
possible order of the matrix elements in Eq.~(\ref{dab}).  Frame
(\textsf{I}) displays additionally orbital configurations
corresponding to this exchange process.} \label{AntiHop}
\end{figure}

The existence of a static DM interaction is not allowed in the
originally proposed tetragonal crystal structure \cite{Yamada89}.
However, AFM resonance at $T = 4.2$~K yields nonvanishing DM vectors
$\textbf{d} \; \|$~[100] and [010] \cite{Yamada94}. The
reexamination of the room temperature crystal structure by x-ray
diffraction \cite{Hidaka98} showed a large standard deviation of the
fluorine ions placed on the magnetic chains away from the $c$ axis.
This was interpreted as a sign of a considerable \emph{static}
displacement of these ions from the $c$ axis, leading to an
orthorhombic distortion allowing for a nonvanishing DM vector
directed along the [110] and equivalent directions. This kind of
displacement in a perovskite structure usually means a rotation of a
CuF$_6$ octahedron as a whole \cite{Deisenhofer02}, but no
displacement of the F$^-$ ions from the $(xy)$ plane has been
detected in Ref.~\onlinecite{Hidaka98}. This fact was emphasized by
Binggeli \etal\cite{Binggeli04} who found in their LDA+U
calculations that this structure relaxes to the conventional
tetragonal structure of KCuF$_3$. This indicates that the
displacements reported in Ref.~\onlinecite{Hidaka98} are rather
\emph{dynamic}, initiated by thermal fluctuations. This assumption
is corroborated by the measurements of the thermal displacement
coefficients of fluorine ions \cite{Buttner90}. They were found to
be very high and exceed strongly those of e.~g. oxygen atoms in an
oxide with the similar crystal structure LaMnO$_3$ \cite{Qiu05}.
Such thermal motions may appear as static on the frequency scale of
x-ray experiments $f \sim 10^{15}$~Hz. In the following we develop a
model of \emph{dynamical anisotropic exchange}, which accounts for
the thermal vibration of an intermediate diamagnetic ion and allows
to explain both the magnitude and the anisotropy of the ESR line in
KCuF$_3$.

We start from the general expression for the antisymmetric exchange
interaction ${\cal H}_{DM} = \mathbf{d}^{(ab)} \cdot [\mathbf{S}_a
\times \mathbf{S}_b]$. In the fifth order of perturbation we have
derived
\begin{eqnarray} \label{dab}
d_{j}^{(ab)} &=&  \frac{2 i}{\Delta_{\eta \xi}} \bigg(
\frac{1}{\Delta_{\kappa \eta} \Delta_{\rho \eta}} +
\frac{1}{\Delta_{\kappa \xi} \Delta_{\rho \xi}} \bigg)  \bigg(
t_{\xi \kappa} t_{\kappa \eta} \cdot \\ \nonumber && \cdot \frac{
\langle \eta \vert \lambda_{a} l_j^{(a)} \vert \zeta
\rangle}{\Delta_{\eta \zeta}} t_{\zeta \rho} t_{\rho \xi} - t_{\eta
\rho} t_{\rho \xi} \frac{ \langle \xi \vert \lambda_{b} l_j^{(b)}
\vert \varphi \rangle}{\Delta_{\xi \varphi}} t_{\varphi \kappa}
t_{\kappa \eta} \bigg),
\end{eqnarray}
where $j = \{x,y,z\}$, $t_{\alpha \beta}$ are the effective hopping
integrals between the states $\vert \alpha \rangle$, $\vert \beta
\rangle$ and a sum over all states of an intermediate ion ($\vert
\kappa \rangle$, $\vert \rho \rangle$) is implied. These virtual
hopping processes are displayed schematically in Fig.~\ref{AntiHop},
where e.~g. frame (\textsf{I}) corresponds to the first term of
Eq.~(\ref{dab}). Note, that expression (\ref{dab}) differs strongly
from the one suggested in Ref.~\onlinecite{Moriya60}, where the
position of the bridging ion, crucial in our case, was not
discussed.

It is well known \cite{Moriya60,Yamada89}, that the static DM
interaction is not allowed in the crystal structure of KCuF$_3$ with
the space group $D_{4h}^{18}$, but now we go beyond the static
configuration. As illustrated in Fig.~\ref{AntiHop}, the
intermediate ion $c$ may be temporarily shifted from its equilibrium
position due to rotation or tilting phonon modes. If the average
time of this displacement $\tau_{\rm F} \approx 2\pi/\omega_{\rm F}$
is large compared to the characteristic time of electron exchange
$\tau_{ab} \approx \hbar /t_{ab}$, several exchange processes shown
in Fig.~\ref{AntiHop} can occur via the displaced ion giving rise to
antisymmetric exchange. Usually $t_{ab} \sim 0.5$~eV what allows to
estimate $\tau_{ab} \approx 10^{-2} \tau_{\rm F}$ using the phonon
frequency $\omega_{\rm F}/2\pi \sim 70$~cm$^{-1}$ as determined from
Raman scattering experiments \cite{Ueda91}.

The relevant pattern of the F$^{-}$ displacements away from the $c$
axis is taken in accord to x-ray diffraction experiments
\cite{Hidaka98}. It can be related to the orbital order in KCuF$_3$
as follows: In Fig.~\ref{AntiHop} we show the geometry of exchange
along with the ground state orbitals $\vert \eta \rangle = \vert
y^2-z^2 \rangle_a$ and $\vert \xi \rangle = \vert x^2-z^2
\rangle_b$. There are four short and two long Cu-F bonds in a
CuF$_6$ octahedron. A displacement of an intermediate F$^-$ ion
along the $y_c$ axis (as shown in the Fig.~\ref{AntiHop})
corresponds to the rotation of the Cu $\vert x^2-z^2 \rangle_b$
orbital and fluorine ion around the $x_b$ axis and does not destroy
short bonds. Note that in this case the $\vert y^2-z^2 \rangle_a$
orbital is fixed by three stationary short bonds and is not rotated
in accordance with the results of the structural analysis
\cite{Hidaka98} which does not reveal any deviations of F$^-$ ions
from the $(ab)$ plane. The second equivalent possibility, a
displacement of the F$^-$ ion along the $x_c$ axis with the
adiabatic rotation of the $\vert y^2-z^2 \rangle_a$ around the $y_a$
axis, is not displayed in the Fig.~\ref{AntiHop}, but will be taken
into account. Substituting the excited state $\vert \varphi \rangle
= \vert yz \rangle_a$ ($\langle d_{yz} \vert l_x \vert
d_{y^{2}-z^{2}} \rangle_a = 2 i$), the Eq.~(\ref{dab}) gives only
one nonzero component of \textbf{d}
\begin{eqnarray} \label{dx}
|d_{x}^{(ab)}| &=& 4 \frac{\lambda_{a}}{\Delta_{cf}} \frac{   (2
t'_{\sigma}+t_{\pi})(t'_{\sigma}+t_{\pi}) t'_{\sigma} }{\Delta_{ab}
\Delta_{ac}^2} \cdot \\ \nonumber && \cdot \big[ -t'_{\sigma}
\textrm{cos}^2(2\theta) + t_{\pi} \textrm{sin}^2(2\theta)  \big]
\textrm{cos}(2\theta) \textrm{sin}(2\theta).
\end{eqnarray}
Note that the factor $\textrm{sin}(2\theta)$ can be rewritten as a
modulus of the vector product $[\mathbf{n}_{ac} \times
\mathbf{n}_{bc}]$, where the unit vectors $\mathbf{n}_{ac}$ and
$\mathbf{n}_{bc}$ connect the spins $a$ and $b$ with the bridging
ion $c$, respectively. Therefore, the rule $\textbf{d}
\parallel [\mathbf{n}_{ac} \times \mathbf{n}_{bc}]$ suggested in
Refs.~\onlinecite{Keffer62,Moskvin77} for the static case is
preserved in our case, too.

The value of $\theta$ is related to the displacement of a F$^-$ ion
from equilibrium position $\textrm{sin}(2\theta)\cdot
R_{\textrm{Cu-F}}$, where $R_{\textrm{Cu-F}}$ is the distance
between the Cu and F ions and can be expressed via phonon operators.
Thus we arrive to a new dynamical form of DM interaction (dDM)
containing simultaneously a phonon and two spin operators. A closely
related form of spin-lattice interaction was introduced by Kochelaev
\cite{Kochelaev99} (see Eq.~(5) in Ref.~\onlinecite{Shengelaya01})
in the context of explanation of the isotope effect on the ESR
linewidth in lightly doped La$_2$CuO$_4$. In our case, however, we
do not need to recall the conventional spin-phonon interaction
(coupling parameter $G$ in
Refs.~\onlinecite{Kochelaev99,Shengelaya01}) at all, which was a
crucial ingredient in that approach.

In the following, we will use this dynamical Hamiltonian to
calculate the ESR linewidth by the method of moments $\Delta H
\propto M_2 (d^2)/J$ \cite{Zakharov08,Eremina03,Zakharov05}, where
upon the thermal averaging over the phonon variables will be
performed. Applying the Einstein model of vibration we get $\langle
\textrm{sin}^2(2\theta) R^2_{\textrm{Cu-F}} \rangle = \frac{2
\hbar}{m \omega_{\rm F}} \textrm{coth}(\hbar \omega_{\rm F} / 2
k_{\rm B} T)$, where $m$ denotes the mass of an oscillating fluorine
ion and $\omega_{\rm F}$ is the frequency of the rotating mode. The
quantity $\langle \textrm{sin}(2\theta) R_{\textrm{Cu-F}} \rangle$
is naturally equal to zero and the static DM interaction does not
exist. To compare with the results of x-ray analysis, it is useful
to introduce the root-mean-square deviation of the F$^-$ ion away
from the $c$ axis $\Delta R_{\perp} = \sqrt{\langle
\textrm{sin}^2(2\theta) R^2_{\textrm{Cu-F}} \rangle}$. Using the
typical rotation frequency of CuF$_6$ octahedra $\omega_{\rm F} \sim
70$~cm$^{-1}$ \cite{Ueda91} one can estimate $\Delta R_{\perp}
\approx 0.07$~{\AA} at $T = 100$~K, what coincides with
Ref.~\onlinecite{Hidaka98}. Therefore, we conclude that these
displacements are rather dynamic than static at $T > T_{\rm N}$.
This conclusion is supported by the latest NQR study
\cite{Mazzoli04} which does not give any evidence for a static
dislocation of F$^-$ ions from the $z$ axis at 77~K as well. To
estimate $d_{x}^{(ab)}$ we will use the values given in
Eq.~(\ref{dx}). Recalling the second possible direction of motion of
intermediate F$^-$ ions, we derive finally
\begin{equation} \label{dEstim}
|d^{(ab)}_x| = |d^{(ab)}_y| \approx 5.6~\textrm{K}.
\end{equation}
Note that this estimate coincides very well with the value
$d^{(ab)}_{\rm AFM} \approx 5.13$~K employed by Yamada and Kato
\cite{Yamada94} to describe the anisotropy of the AFM resonance line
below $T_{\rm N}$. Moreover, it yields the correct linewidth of
about $10^3$~Oe in the high-temperature limit.

The anisotropic interchain exchange can be neglected, because of the
small constant of isotropic exchange $J_\perp \ll J$ and the
orthogonality of orbital states. The dynamical form of SAE would
arise in the higher order of perturbation theory as compared to the
dDM interaction and is expected to be negligibly small. The
anisotropic Zeeman effect does not contribute, because we do not
observe any significant field dependence of the linewidth up to the
frequency of 150~GHz (Fig.~\ref{Exp}). Therefore, the spin
relaxation mechanisms described above, the static SAE
(Eq.~\ref{Destim}) and the dDM interaction (Eq.~\ref{dEstim})
between the Cu ions along the $c$ axis are the only relevant sources
of the ESR line broadening in KCuF$_3$. Indeed, these two broadening
mechanisms can explain both the magnitude and the anisotropy of the
ESR linewidth. Figure~\ref{Exp}(b) shows a fit of the angular
dependence of $\Delta H$ at room temperature and emphasizes the
contributions of SAE and dDM interaction to the linewidth, which
reflects clearly the dominant role of the dDM interaction. This
indicates that KCuF$_3$, despite its paradigmatic status for orbital
order, is governed by strong fluctuations in the lattice and orbital
sector.

In summary, we have introduced a dynamical Dzyaloshinsky-Moriya
(dDM) interaction which allowed to explain the long standing mystery
about the origin of the huge ESR linewidth in KCuF$_3$ at $T>T_{\rm
N}$ in consistence with other experimental findings about the
properties of this quasi-1D system. Such a dDM interaction becomes
effective, if the characteristic time of the dynamic distortion
resulting in a nonzero DM vector is large compared to the time scale
of the exchange interaction and if the amplitude of these
distortions is high. This is the case for low-lying optical modes
with the tendency to soften to low temperatures. Therefore, such
kind of interaction may be on general useful for understanding the
spin dynamics of other materials with soft-mode vibrations.

We thank I.~Leonov for useful discussions. This work was supported
by the DFG within SFB 484 (Augsburg) and by the RFBR (Grant No.
06-02-17401-a). One of us (D.~V.~Z.) was supported by VW-Stiftung.



\begin{thebibliography}{99}

\item[*] Corresponding author: \\Dmitri.Zakharov@physik.uni-augsburg.de

\bibitem{Kugel82} E.~Dagotto \textit{et al.}, Phys. Rep. {\bf 344}, 1 (2001);
K.I. Kugel, D.I. Khomskii, Usp. Fiz. Nauk {\bf 136}, 621
(1982).

\bibitem{Lake05} B.~Lake \textit{et al.}, Nature Materials {\bf 4}, 329 (2005).

\bibitem{Hidaka98} M.~Hidaka \textit{et al.}, J.~Phys.~Soc.~Jpn. {\bf 67}, 2488 (1998).

\bibitem{Yamada89} I.~Yamada \textit{et al.}, J.~Phys.:~Cond.~Mat. {\bf 1}, 3397 (1989).

\bibitem{Yamada94} I.~Yamada, N.~Kato, J.~Phys.~Soc.~Jpn. {\bf 63}, 289 (1994).

\bibitem{Mazzoli04} C.~Mazzoli \textit{et al.}, J.~Magn.~Magn.~Mat. \textbf{242},
935 (2002).

\bibitem{Binggeli04} N.~Binggeli, M.~Altarelli,
Phys.~Rev.~B {\bf 70}, 085117 (2004).


\bibitem{Caciuffo02} R.~Caciuffo \textit{et al.}, Phys.~Rev.~B \textbf{65}, 174425
(2002).

\bibitem{Ivannikov02} D. Ivannikov \textit{et al.}, Phys. Rev. B \textbf{65}, 214422 (2002).

\bibitem{Ishii90} T.~Ishii, I.~Yamada, J.~Phys.:~Cond.~Mat. {\bf 2}, 5771 (1990).

\bibitem{Ikebe71} M.~Ikebe, M.~Date, J.~Phys.~Soc.~Jpn. {\bf 30}, 93
(1971).

\bibitem{Zakharov08} D.V. Zakharov \textit{et al.}, in \textit{Quantum
Magnetism}, edited by B.~Barbara \textit{et al.}, (Springer, The
Netherlands 2008).

\bibitem{Oshikawa02} M.~Oshikawa, I.~Affleck, Phys. Rev. B {\bf 65}, 134410 (2002).

\bibitem{Ivanshin03} V.A.~Ivanshin \textit{et al.}, Phys.~Rev.~B {\bf 68}, 064404 (2003).

\bibitem{KrugvNidda02} H.-A.~Krug~von~Nidda \textit{et al.}, Phys. Rev. B {\bf 65}, 134445 (2002).

\bibitem{Eremina03} R.M.~Eremina \textit{et al.}, Phys. Rev. B {\bf 68}, 014417 (2003).

\bibitem{Eremin05} M.V.~Eremin \textit{et al.}, Phys.~Rev.~Lett. {\bf 96}, 027209 (2006).

\bibitem{Zakharov05} D.V. Zakharov \textit{et al.}, Phys.~Rev.~B
\textbf{73}, 094452 (2006).

\bibitem{Satija80} S.K.~Satija \textit{et al.}, Phys.~Rev.~B {\bf 21}, 2001 (1980).

\bibitem{Deisenhofer02} J. Deisenhofer \textit{et al.}, Phys. Rev. B {\bf 65},
104440 (2002).

\bibitem{Buttner90} R.H. Buttner \textit{et al.}, Acta Cryst. {\bf B46}, 131 (1990).

\bibitem{Qiu05} X.~Qiu \textit{et al.}, Phys.~Rev.~Lett. {\bf 94},
177203 (2005).

\bibitem{Moriya60} T.~Moriya, Phys. Rev. {\bf 120}, 91 (1960).

\bibitem{Ueda91} T.~Ueda \textit{et al.}, Solid~State~Comm. {\bf 80}, 801 (1991).

\bibitem{Moskvin77} A.S.~Moskvin, I.G.~Bostrem, Fiz. Tverd. Tela {\bf 19}, 1616
(1977) [Sov.~Phys.~Solid~State \textbf{73}, 1532 (1977)].

\bibitem{Keffer62} F. Keffer, Phys. Rev. {\bf 126}, 896 (1962).

\bibitem{Kochelaev99} B. I. Kochelaev, J. Supercond. {\bf 12}, 53
(1999).

\bibitem{Shengelaya01} A. Shengelaya \textit{et al.}, Phys. Rev. B
{\bf 63}, 144513 (2001).

\end{thebibliography}
\end{document}